# Topological non-Hermitian origin of surface Maxwell waves

Konstantin Y. Bliokh[1,2], Daniel Leykam[3], Max Lein[4], and Franco Nori[1,5]

[1]*Theoretical Quantum Physics Laboratory, RIKEN Cluster for Pioneering Research, Wako-shi, Saitama 351-0198, Japan*
[2]*Nonlinear Physics Centre, RSPE, The Australian National University, ACT 0200, Australia*
[3]*Center for Theoretical Physics of Complex Systems, Institute for Basic Science (IBS), Daejeon 34126, Republic of Korea*
[4]*Advanced Institute of Materials Research, Tohoku University, Sendai 980-8577, Japan*
[5]*Physics Department, University of Michigan, Ann Arbor, Michigan 48109-1040, USA*

**Maxwell electromagnetism, describing the wave properties of light, was formulated 150 years ago. More than 60 years ago it was shown that interfaces between optical media (including dielectrics, metals, negative-index materials) can support *surface electromagnetic waves*, which now play crucial roles in plasmonics, metamaterials, and nano-photonics. Here we show that surface Maxwell waves at interfaces between homogeneous, isotropic media described by real permittivities and permeabilities have a purely *topological* origin explained by the bulk-boundary correspondence. Importantly, the topological classification is determined by the *helicity* operator, which is generically *non-Hermitian* even in lossless optical media. The corresponding topological invariant, which determines the number of surface modes, is a $\mathbb{Z}_4$ number (or a pair of $\mathbb{Z}_2$ numbers) describing the winding of the complex helicity spectrum across the interface. Moreover, there is an additional pair of non-topological $\mathbb{Z}_2$ indices, which describe zones of the TE and TM polarizations at the phase diagram of surface modes. Our theory provides a new twist and insights for several areas of wave physics: Maxwell electromagnetism, topological quantum states, non-Hermitian wave physics, and metamaterials.**

## 1. Introduction

Classical and quantum waves underlie the most fundamental entities in nature: light, sound, fields, and matter. Recently, an important role of *topology* in wave systems was revealed, describing the appearance of surface waves at interfaces between topologically-different media [1,2]. This brought about the explanation of various physical phenomena (e.g., the quantum Hall effect [3,4]), the prediction of new phenomena (e.g., topological insulators [1,2]), and eventually resulted in the Nobel Prize in physics in 2016. While it was initially believed that topological effects are particular to quantum systems, they are universal *wave* phenomena, which since then have been realized in a wide range of classical waves, including electromagnetic [5–7], acoustic [8], mechanical [9], and hydrodynamic [10] systems.

Optics and electromagnetism provide one of the best platforms for studying fundamental relativistic wave phenomena, because classical Maxwell equations represent relativistic wave equations for massless spin-1 particles, i.e., photons within the first-quantization approach [11–13]. (This explains the mathematical similarities to the Dirac equation, even though Maxwell equations describe *classical* electromagnetic fields.) Moreover, studies of *surface electromagnetic waves* at interfaces between different media resulted in the rapid development of several areas of modern photonics, such as plasmonics [14,15] and negative-index metamaterials [16–18]. Not surprisingly, the discovery of topological wave phenomena generated the rapidly developing field of topological photonics [19,20].

Topological electromagnetic modes have been predicted and demonstrated in rather complicated nanostructured metamaterials, which mimic condensed-matter crystals with topologically-nontrivial electron Hamiltonians. This approach requires considerable engineering efforts and suffers from inevitable losses, imperfections, etc. In contrast, in this paper, we reveal nontrivial topological properties for the most basic form of Maxwell equations involving only isotropic lossless homogeneous media characterized by the permittivity $\varepsilon$ and permeability $\mu$.

We show that all surface Maxwell waves appearing at interfaces between media with different signs of $\varepsilon$ and $\mu$ are *topological* in nature. Here the term "topological" is justified in two ways. First, we describe the *bulk-boundary correspondence*, where the number of surface modes is determined by the contrast of a topological bulk invariant across the interface [1,2,19,20]. Importantly, this bulk invariant originates from the *helicity operator* of photons in a medium. This is the central difference of our work as compared with previously described topological systems based on the *Hamiltonian* operator. Furthermore, this helicity operator is generically *non-Hermitian* [21,22] and has purely imaginary eigenvalues in "metallic" media with $\varepsilon\mu < 0$ [23]. The topological bulk invariant is a $\mathbb{Z}_4$ number (or a pair of $\mathbb{Z}_2$ numbers), which describe the *phase* of the gapped helicity spectrum in a medium. The winding of this spectrum across the interface exactly corresponds to the number of surface electromagnetic modes which are *zero-helicity* TE or TM-polarized waves. Second, we connect the topology of the bulk system to the topology of the parameter $(\varepsilon,\mu)$ space; this is analogous to earlier works [24,25] in the condensed-matter context. For Maxwell waves, the parameter space is split into four simply connected quadrants excluding the $\varepsilon = 0$ and $\mu = 0$ lines, where the helicity is ill-defined. The helicity-based topological bulk invariant labels these quadrants of the parameter space. In addition to the topological invariant that provides the number of surface modes, we introduce a pair of non-topological $\mathbb{Z}_2$ indices which separate the zones of the TE and TM polarizations of surface modes in the phase diagram of surface modes.

Our non-Hermitian topological theory allows us to fully explain the nontrivial phase diagram of Maxwell surface modes, which includes well-known examples of surface plasmon-polaritons at metal-dielectric and negative-index interfaces, and to augment it with previously-overlooked *evanescent surface waves* decaying along the propagation direction or/and in time. Although this diagram can be obtained from the standard Maxwell equations and boundary conditions, only the present topological theory explains *why* surface Maxwell modes of different TE and TM polarizations exist in the corresponding regions of the parameter $(\varepsilon,\mu)$-space.

## 2. Results

### 2.1. Winding of the helicity spectrum of photons in a medium

We start with the simplest example of topological surface modes, namely, the Jackiw-Rebbi edge states in the Dirac equation [1,2,26]. The bulk spectrum $E(\mathbf{p}) = \pm\sqrt{p^2 + m^2}$ of the Dirac equation is characterized by the *energy gap* $2m$ (we use $\hbar = c = 1$ units) determined by the mass $m$. Then, an interface between two media with opposite masses $m_1 = -m_2 \equiv m$ supports a topological surface state with *massless* spectrum $E^{\text{surf}} = \pm p^{\text{surf}}$, Fig. 1(a). This edge mode is protected by the difference of the $\mathbb{Z}_2$ *topological winding number* $w = \frac{1}{2}\text{sgn}(m)$ in the two media [1,2,26]. The transition between the two media can be viewed as a $\pi$ rotation (i.e., winding) of the rest energies $E_0 \equiv E(0) = \pm m \to \pm e^{i\pi} m = \mp m$ in the complex-energy (mass) plane, Fig. 2(a), which illuminates the Möbius-strip-like $\mathbb{Z}_2$ topology.

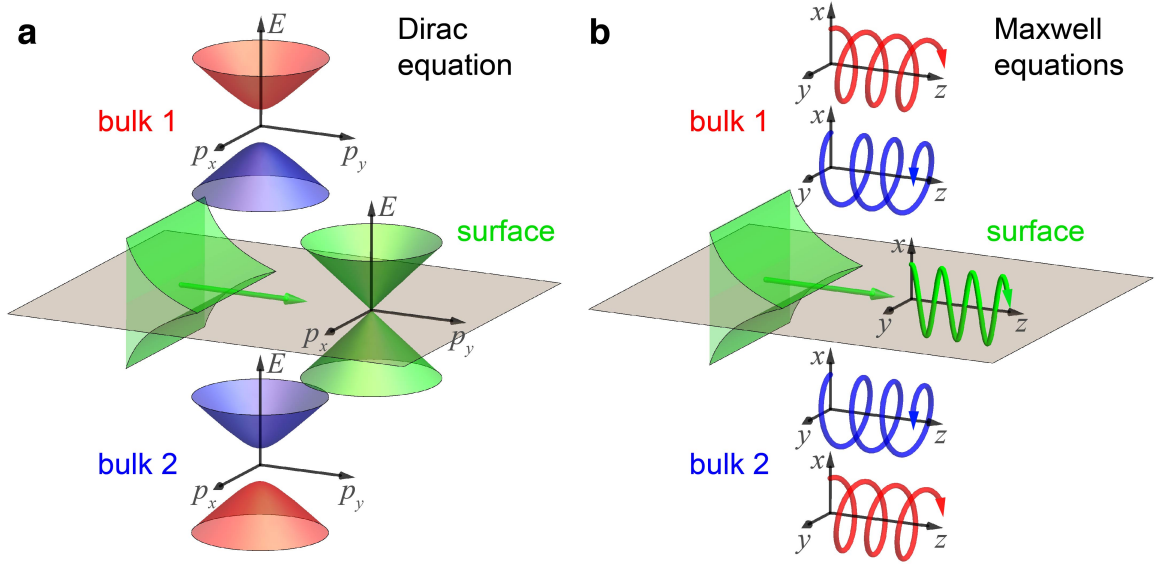

**Fig. 1. Schematics of topological surface modes in the Dirac and Maxwell equations. (a)** The Dirac equation with a finite *mass* $m$ is characterized by the gapped bulk spectrum $E(\mathbf{p})$. An interface between "media" with opposite-sign masses $\pm m$, and bulk spectra (schematically shown in red and blue), supports topological surface modes with *massless* spectrum (shown in green) [1,2,26]. **(b)** Maxwell equations possess massless bulk spectra (not shown here), which are double-degenerate with respect to opposite *helicity* states. These bulk helicity eigenmodes have opposite circular polarizations, i.e., chiral spatial distributions of the electric and magnetic fields (shown in red and blue here). An interface between two media with different helicity properties (controlled by the signs of the permittivity $\varepsilon$ and permeability $\mu$ of the medium) supports *zero-helicity* surface waves with TE or TM linear polarizations (shown in green) [14–18].

Consider now electromagnetic waves (photons) described by the source-free Maxwell equations. Photons do not have mass but they possess another fundamental property: *helicity*, which can be associated with the projection of photon's spin $\mathbf{S}$ onto the direction of its momentum: $\mathfrak{S} = \mathbf{S} \cdot \mathbf{p} / |\mathbf{p}|$ [23,27–30]. It is known that the helicity of free-space photons has two eigenvalues $\sigma = \pm 1$, corresponding to the right-hand and left-hand circularly polarized waves, Fig. 1(b), whereas the independent zero-helicity state is forbidden because of the transversality of electromagnetic waves [27]. Thus, one can say that Maxwell bulk eigenmodes are characterized by the *helicity gap*, Fig. 2(b).

In this paper, we deal with Maxwell waves in isotropic lossless media characterized by a real-valued permittivity $\varepsilon$ and permeability $\mu$. The possible dispersion of these parameters does not affect our considerations and is neglected hereafter. We will also use the refractive index $n$ and dimensionless impedance $Z$ of the medium, with $|n| = \sqrt{|\varepsilon\mu|}$, $|Z| = \sqrt{|\mu/\varepsilon|}$, and the signs defined as shown in Fig. 2(c) for four possible types of media [17]. The opposite refractive-index signs in the "positive-index" and "negative-index" materials reflect the fact that the complex energy flux (Poynting vector) and momentum (wavevector) are parallel and anti-parallel in such media (see Supplementary Materials) [16,17]. The gapless bulk energy spectrum of electromagnetic waves is determined by the dispersion relation: $\omega^2 = k^2 / (\varepsilon\mu)$ ($\omega$ is the frequency, **k** is the wavevector), so that the bulk modes are propagating in transparent media with $\varepsilon\mu > 0$, and become purely evanescent, with imaginary wavevector or frequency, in "metals" with $\varepsilon\mu < 0$.

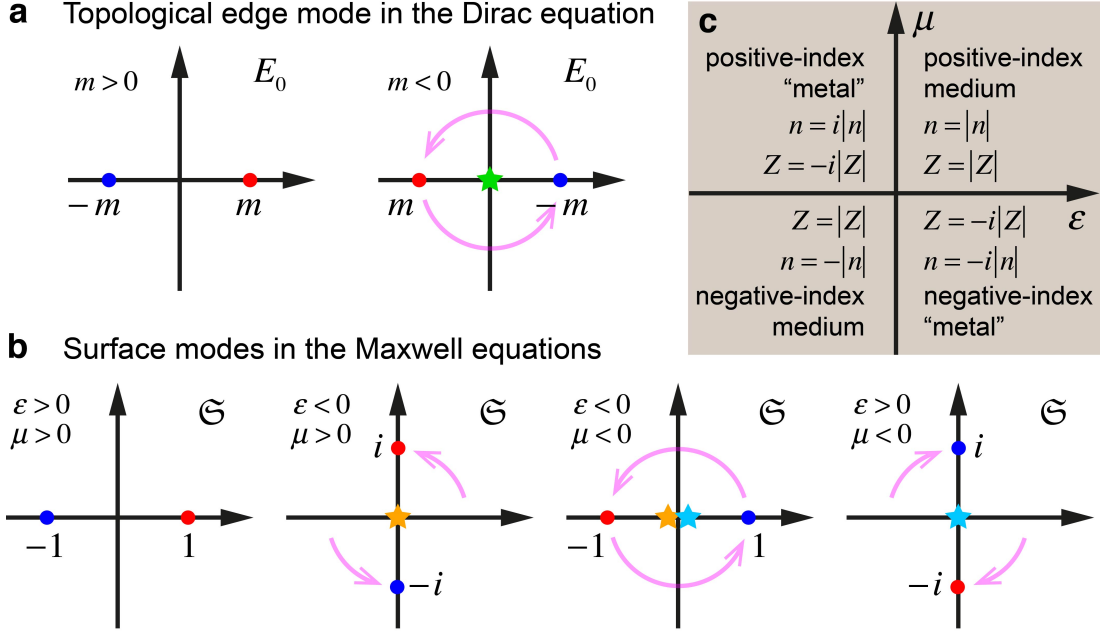

**Fig. 2. Winding of the energy and helicity spectra and the appearance of surface modes in the Dirac and Maxwell equations.** **(a)** Changing the sign of the mass $m$ in the Dirac equation is equivalent to a $\pi$ rotation in the complex-mass (rest-energy $E_0$) plane, which results in a single zero-mass surface mode (shown by the star symbol) protected by the topological $\mathbb{Z}_2$ winding number [1,2,26]. **(b)** Changing the signs of the permittivity $\varepsilon$ and permeability $\mu$ in Maxwell equations produces $\pm\pi/2$ and $\pi$ rotations in the complex helicity ($\mathfrak{S}$) plane, Eq. (2). This results in the appearance of one or two zero-helicity (TE and TM) surface modes [14–18,33–35] (shown by the star symbols) described by the topological $\mathbb{Z}_4$ number (3). **(c)** The "medium-index diagram" showing the signs of the refractive index $n$ and impedance $Z$ in four possible types of media (see Supplementary Materials).

Maxwell equations for monochromatic light in a medium can be written in a quantum-like form as a Weyl-type equation [11–13,23,30,31]:

$$\left(\hat{\mathbf{S}}\cdot\hat{\mathbf{p}}\right)\psi = -\omega\hat{\sigma}^{(m)}\psi, \quad \hat{\sigma}^{(m)} = \begin{pmatrix} 0 & -i\mu \\ i\varepsilon & 0 \end{pmatrix}, \quad \psi = \begin{pmatrix} \mathbf{E} \\ \mathbf{H} \end{pmatrix}. \quad (1)$$

Here, $\psi$ is the 6-component "wavefunction", $\hat{\mathbf{p}} = -i\nabla$ is the momentum operator, $\hat{\mathbf{S}}$ is the vector of $3\times 3$ spin-1 matrices, acting on the Cartesian components of the fields as $\hat{\mathbf{S}}\cdot\hat{\mathbf{p}} = \nabla\times$, whereas the matrix $\hat{\sigma}^{(m)}$, describing the properties of the medium, acts on the "electric-magnetic" degrees of freedom, i.e., intermixes the **E** and **H** fields. The presence of a medium modifies the scalar product in this quantum-like approach, so that $\langle\psi|\psi\rangle = \tilde{\psi}^\dagger\cdot\psi$ with the adjoint "left" vector being $\tilde{\psi} = (\varepsilon\mathbf{E},\mu\mathbf{H})^T \equiv (\mathbf{D},\mathbf{B})^T$ [23,32]. Using this formalism, it was recently shown that the helicity remains a fundamental physical property of electromagnetic waves in isotropic dispersive media [22,23]. Consider circularly-polarized plane waves $\psi^{(\sigma)}$ with the electric field $\mathbf{E}^{(\sigma)} = (1, i\sigma, 0)\exp(i\mathbf{k}\cdot\mathbf{r} - i\omega t)$ ($\sigma = \pm 1$ determines the sign of the circular polarization), and the corresponding magnetic field $\mathbf{H}^{(\sigma)} = -i\sigma Z^{-1}\mathbf{E}$ (see Supplementary Materials). These are eigenmodes of the helicity operator in the medium, $\hat{\mathfrak{S}}$ [23], $\hat{\mathfrak{S}}\psi^{(\sigma)} = \mathfrak{S}\psi^{(\sigma)}$, with *complex eigenvalues* as follows:

$$\hat{\mathfrak{S}} = -\frac{\hat{\sigma}^{(m)}}{|n|} = \begin{pmatrix} 0 & i\eta Z \\ -i\eta Z^{-1} & 0 \end{pmatrix}, \quad \mathfrak{S} = \eta\sigma. \qquad (2)$$

Here, $\eta = n/|n|$ indicates the phase of the refractive index, and we note that imaginary helicity makes physical sense because the canonical momentum (wavevector) becomes imaginary (while the spin remains real) in metallic media [23]. Remarkably, the helicity $\mathfrak{S}$ always equals 1 in absolute value, but its *phase* essentially depends on the signs of $\varepsilon$ and $\mu$, i.e., is different in the four types of optical media mentioned above. At the dividing lines $\varepsilon = 0$ and $\mu = 0$, separating different phases, the helicity is ill-defined (as well as the diverging energy eigenvalue $\omega$).

Thus, the "helicity gap" is always present in optical media (apart from the singular $\varepsilon = 0$ and $\mu = 0$ cases), whereas the media with different signs of $(\varepsilon, \mu)$ are related by $\pi/2$, $\pi$, and $-\pi/2$ rotations in the complex helicity plane, as shown in Fig. 2(b). This suggests that electromagnetic media are split into four topologically different classes, described by the *topological bulk invariant*, which is a $\mathbb{Z}_4$ number or, equivalently, a pair of $\mathbb{Z}_2$ numbers:

$$w(\varepsilon,\mu) = \frac{2}{\pi}\mathrm{Arg}\left[\eta(\varepsilon,\mu)\right] \quad \text{or} \quad \{w^{\mathrm{TM}}(\varepsilon,\mu), w^{\mathrm{TE}}(\varepsilon,\mu)\} = \frac{1}{2}\{1-\mathrm{sgn}(\varepsilon), 1-\mathrm{sgn}(\mu)\}. \qquad (3)$$

Here the $\mathbb{Z}_4$ number $w$ takes on values $0, \pm 1, 2$ in four types of media shown in Fig. 2(c), while the $\mathbb{Z}_2$ numbers $w^{\mathrm{TM,TE}}$ take on values $0, 1$. We will refer to the invariant $w$ as to the *helicity winding number*, because the contrast of this invariant between two optical media describe the winding of the complex helicity spectrum across the interface.

Most importantly, interfaces between different media indeed support surface electromagnetic modes [14–18,33–35], which are in agreement with the differences of the topological numbers (3) across the interface. First, the surface Maxwell modes always have *zero helicity*, $\mathfrak{S}^{\mathrm{surf}} \equiv 0$, similarly to the zero-mass modes in topological insulators [1,2,26], Fig. 1. Indeed, the surface Maxwell waves are either transverse-electric (TE) or transverse-magnetic (TM), so that the product of the magnetic and electric wave fields, that determines the expectation value of the helicity operator (2), vanishes identically: $\mathfrak{S} \propto \mathbf{H}^* \cdot \mathbf{E} \equiv 0$ [23,31] (in agreement with this, the spin of these modes is orthogonal to the wavevector: $\mathbf{S} \cdot \mathbf{k} = 0$ [35]). Second, the number of TE and TM surface modes at the interface is exactly determined by the differences of the topological numbers (3):

$$N_{\mathrm{surf}} = \left|w(\varepsilon_2,\mu_2) - w(\varepsilon_1,\mu_1)\right| = \left|w(\varepsilon_r,\mu_r)\right| = N_{\mathrm{surf}}^{\mathrm{TE}} + N_{\mathrm{surf}}^{\mathrm{TM}},$$

$$N_{\mathrm{surf}}^{\mathrm{TM,TE}} = \left|w^{\mathrm{TM,TE}}(\varepsilon_2,\mu_2) - w^{\mathrm{TM,TE}}(\varepsilon_1,\mu_1)\right| = \left|w^{\mathrm{TM,TE}}(\varepsilon_r,\mu_r)\right|. \qquad (4)$$

where the subscripts "1", "2", and "r" indicate the parameters of the two bulk media and the relative parameters characterizing the interface: $(\varepsilon_r, \mu_r) = (\varepsilon_2/\varepsilon_1, \mu_2/\mu_1)$. Note that in the first Eq. (4) the difference should be considered within the cyclic $\mathbb{Z}_4$ group: e.g. $2-(-1) = -1$ rather than $3$, because the helicity spectra of the corresponding media are related by the $-\pi/2$ rather than $3\pi/2$ rotation. Equations (4) determine the *bulk-boundary correspondences* for the topological numbers (3) and surface Maxwell waves. In simple words, Eqs. (4) state that a single TM (TE) surface mode exists at an interface where only the permittivity $\varepsilon$ (permeability $\mu$) changes its sign, and two surface modes (TE and TM) exist at interfaces where both $\varepsilon$ and $\mu$ change sign. This is shown in the phase diagram in Fig. 3(a) and is in perfect agreement with the properties of surface Maxwell waves known in plasmonics and metamaterials [14–18].

Remarkably, the helicity winding number (3) can also be associated with the phase of the topological *Chern number* of photons [35], which is also intimately related to the helicity and can become *complex* in "metallic" media that only support evanescent modes. In free space ($\varepsilon = \mu = 1$), the Berry curvature for photons is a monopole of charge $\sigma$ at the origin of the momentum space: $\mathbf{F}^\sigma = \sigma \mathbf{k}/k^3$. Integrating it over momentum-space sphere yields the helicity-dependent Chern number $C^\sigma = 2\sigma$ [35]. Extending this construction to isotropic media, we find that the momentum space becomes *complex* (assuming real frequency $\omega$, the wavevectors $\mathbf{k}$ become imaginary in metallic media with $\varepsilon\mu < 0$). This results in the substitution $\mathbf{k}/k \to \eta \mathbf{k}/k$, and the Chern number becomes $C^\sigma = 2\eta\sigma = 2\mathfrak{S}$ (see Supplementary Materials). Thus, transitions between media with different signs of $\varepsilon$ or $\mu$ are accompanied by discrete changes of the *phase* of the complex Chern numbers, and the topological number (3) is determined by the phase of the spin Chern number: $w = (2/\pi)\mathrm{Arg}(\sigma C^\sigma)$. This illuminates the topological helicity properties of Maxwell equations in media and shows that these are quite different as compared to Hermitian topological insulators with gapped energy spectra and real Chern numbers.

## 2.2. Non-Hermitian properties of the helicity and Maxwell equations

The above consideration reveals another fundamental peculiarity of the helicity-based description of photons in a medium. Namely, the helicity operator (2) is essentially *non-Hermitian*, as it is clearly seen from its purely imaginary spectrum in metallic media with $\varepsilon\mu < 0$. Therefore, the corresponding helicity-based form of Maxwell equations, Eq. (1), is also effectively non-Hermitian. Indeed, expanding the matrix $\hat{\sigma}^{(m)} = -|n|\hat{\mathfrak{S}}$ in terms of the Hermitian Pauli matrices $\hat{\sigma}_i$, we write Maxwell equations as:

$$(\hat{\mathbf{S}}\cdot\hat{\mathbf{p}})\psi = -\frac{\omega}{2}\left[(\varepsilon+\mu)\hat{\sigma}_2 + i(\varepsilon-\mu)\hat{\sigma}_1\right]\psi. \tag{5}$$

Despite the non-Hermiticity of the operator in the right-hand side of this equation, its spectrum can be real (in transparent media with $\varepsilon\mu > 0$) because there is time-reversal symmetry $\hat{K}\hat{\sigma}_3$, where $\hat{K}$ is the complex conjugation [21,22,36]. Notably, it is known that Maxwell equations in a medium can be treated as a Hermitian *energy* eigenvalue problem, i.e., the frequency $\omega$ can be always chosen to be real (with the wavevector $\mathbf{k}$ becoming imaginary in metallic media) [5,23,32]. However, an important fact is missing in this Hermitian consideration with the modified inner product $\langle\psi|\psi\rangle = \tilde{\psi}^\dagger\cdot\psi$: it is valid for arbitrary $(\varepsilon,\mu)$ *apart from the $\varepsilon = 0$ and $\mu = 0$ values*. The energy eigenvalues diverge at these values, $\omega \to \infty$, while the inner product coefficients vanish. Remarkably, these singular $\varepsilon = 0$ and $\mu = 0$ values correspond to *exceptional points* [22,37] of the operator $\hat{\sigma}^{(m)}$ in the helicity-based form (1) and (5) of Maxwell equations. The bulk helicity spectrum changes from real ($\varepsilon\mu > 0$) to imaginary ($\varepsilon\mu < 0$) at these points. Moreover, in each of the exceptional points, the bulk modes ($\psi = (\mathbf{E},\mathbf{H})^T \propto (1,-i\sigma)^T$ are the eigenstates of $\hat{\sigma}_2$ in the vacuum) tend to a *single "chiral" mode* [37–39] (the eigenstate of $\hat{\sigma}_3$): $\psi_c \propto (1,0)^T$ or $\psi_c \propto (0,1)^T$, having only an electric or magnetic field (see Supplementary Materials). These "chiral" modes play crucial roles in the "epsilon-near-zero" or "mu-near-zero" materials [40].

It is known in the theory of non-Hermitian systems that exceptional points are spectral degeneracies with nontrivial topological structure [22,37,41,42]. Namely, they have the topology of branch points, and it is impossible to introduce an unambiguous global labeling of eigenvalues in the vicinity of exceptional points. Thus, the parameter $(\varepsilon, \mu)$ space of Maxwell equations is actually split into four simply connected domains (quadrants) separated by the "exceptional lines" $\varepsilon = 0$ and $\mu = 0$. The effective Hermitian description [5,23,32] is possible in each of these domains but not globally over the whole $(\varepsilon, \mu)$ space. The helicity winding number (3) labels these topologically different quadrants of the parameter space, and has essentially non-Hermitian origin.

### 2.3. Additional polarization indices

As mentioned above, the quantum-like formalism for Maxwell equations (1) determines the biorthogonal set of "right" and "left" eigenvectors $\psi$ and $\tilde{\psi}$ [23]. However, this choice is not unique. Alternatively, Maxwell equations can be formulated for the vectors $\psi' = (\mathbf{E}, \mathbf{B})^T$ and $\tilde{\psi}' = (\mathbf{D}, \mathbf{H})^T$. In this case, Eq. (3) becomes:

$$(\hat{\mathbf{S}} \cdot \hat{\mathbf{p}})\psi' = -\frac{\omega}{2}\left[(\varepsilon\mu + 1)\hat{\sigma}_2 + i(\varepsilon\mu - 1)\hat{\sigma}_1\right]\psi'. \qquad (6)$$

The non-Hermitian operator in the right-hand side of Eq. (6) has the same exceptional points as in Eq. (5). However, the Hermitian and non-Hermitian parts of the operators in Eqs. (3) and (4) differ from each other. Recently, analyzing topological edge modes in non-Hermitian quantum systems [43], we showed that the sign of the non-Hermitian part of the operator can play an important role in this problem. For the operators in Eqs. (5) and (6), this results in *a pair of $\mathbb{Z}_2$ indices*:

$$v(\varepsilon, \mu) = \frac{1}{2}\{\text{sgn}(\varepsilon - \mu), \text{sgn}(\varepsilon\mu - 1)\}. \qquad (7)$$

As we show below, these indices describe the polarization TE/TM properties of surface modes, and therefore we will refer to these as to "*polarization indices*". Importantly, for a *single* medium, one can scale the electric and magnetic fields such that this will remove the non-Hermitian $\hat{\sigma}_1$-term in Eq. (5) or (6). In particular, scaling $\psi = (\alpha\mathbf{E}, \beta\mathbf{H})^T$ with $\beta/\alpha = Z$ yields $-n\omega\hat{\sigma}_2\psi$ in the right-hand side of Eq. (5). However, such scaling is singular at the exceptional points $\varepsilon = 0$ and $\mu = 0$, and, furthermore, it cannot remove the $\hat{\sigma}_1$-term simultaneously in *two* media. Applying the above scaling to the first medium, $\alpha/\beta = Z_1$, we find that the Maxwell equations in the second medium are given by Eqs. (5) and (6) with the substitution

$$(\varepsilon_2, \mu_2) \to (\varepsilon_r, \mu_r), \qquad \omega \to n_1\omega. \qquad (8)$$

Thus, the fundamental interface properties and surface modes must depend on the polarization indices (7) involving the *relative* permittivity and permeability: $v(\varepsilon_r, \mu_r)$. In contrast to the topological numbers (3) and the bulk-boundary Eq. (4), the polarization indices of the relative interface parameters, $v(\varepsilon_r, \mu_r)$, cannot be expressed via differences of the corresponding bulk indices $v(\varepsilon_1,\mu_1)$ and $v(\varepsilon_2,\mu_2)$. This shows that the polarization indices (7) are *not* topological numbers, and there is no bulk-boundary correspondence for these. The role of these indices is revealed below.

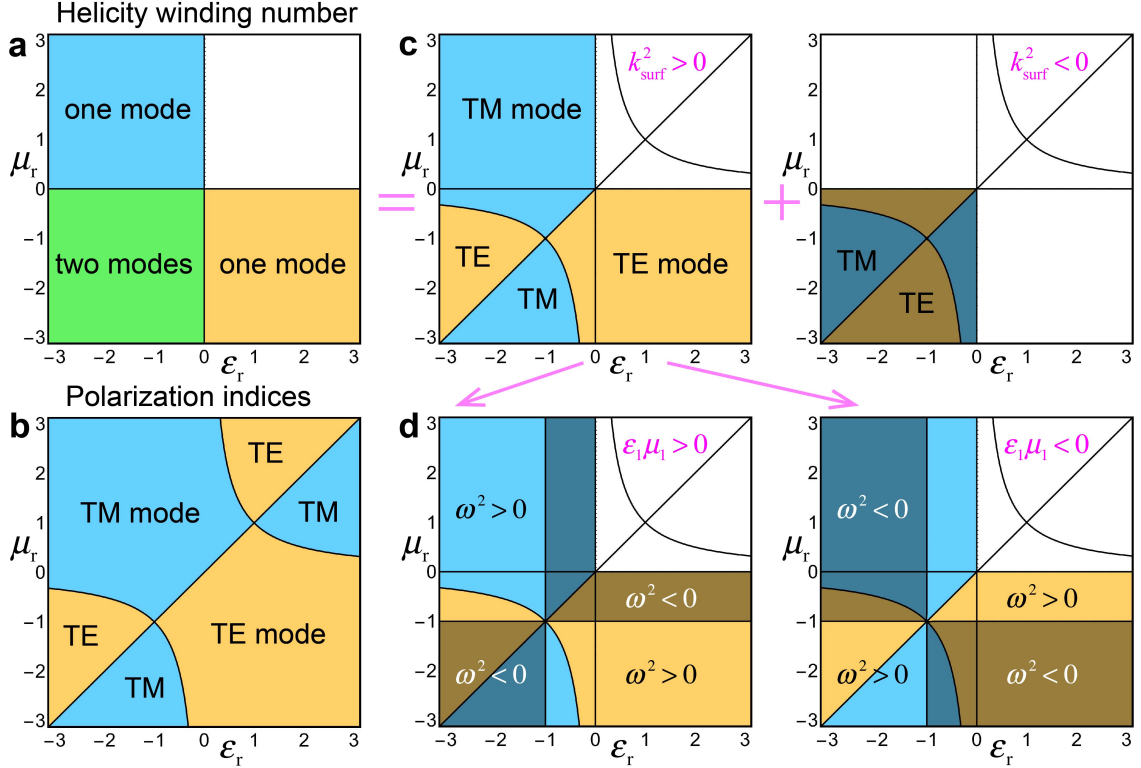

**Fig. 3. Phase diagrams of surface Maxwell waves. (a)** Zones of the existence of zero, one, and two surface zero-helicity modes described by the topological "helicity winding number" (3) and the bulk-boundary correspondence (4), see Fig. 2(b). **(b)** Phase separation of the TE and TM modes described by the "polarization indices" (7). **(c)** The phase diagrams resulting from the combination of **(a)** and **(b)**. The "two-mode" quadrant $(\varepsilon_r<0,\mu_r<0)$ has *both* TE and TM modes in every point, but only *one* of these is propagating (i.e., having real wavevector $k_{\text{surf}}$), while the other one is evanescent (having imaginary $k_{\text{surf}}$). **(d)** Splitting the phase diagram **(c)** with real $k_{\text{surf}}$ into zones with real (bright areas) and imaginary (dark areas) frequency $\omega_{\text{surf}}$. These zones swap upon the inversion of the sign of the squared refractive index of the first medium, $n_1^2 = \varepsilon_1\mu_1$.

## 2.4. Phase diagrams for surface Maxwell waves

The detailed phase diagram of surface Maxwell modes can now be constructed using the topological invariants (3) with the bulk-boundary correspondence (4), augmented by the polarization indices (7) and simple symmetry arguments. First, as it was mentioned above, the helicity winding number (3), $w(\varepsilon_r,\mu_r)$, yields a diagram in the $(\varepsilon_r,\mu_r)$-plane, Fig. 3(a), which determines the *number of surface modes* according to Eq. (4). These modes must have vanishing helicity: $\mathfrak{S}^{\text{surf}} \propto \mathbf{H}^* \cdot \mathbf{E} \equiv 0$. Taking into account the symmetry of a planar interface between two isotropic media implies that the surface modes must be TE or TM polarized, i.e., having electric and magnetic fields parallel and perpendicular to the interface and orthogonal to each other.

Second, the indices $v(\varepsilon_r, \mu_r) \equiv \{v_1, v_2\}$ determine the *separation between the TE and TM phases*. Indeed, from the dual symmetry between the electric and magnetic quantities ($\varepsilon_r \leftrightarrow \mu_r$, TE$\leftrightarrow$TM) and the spatial inversion symmetry, which exchanges the two media, $1 \leftrightarrow 2$, and produces the substitution $(\varepsilon_r, \mu_r) \rightarrow (1/\varepsilon_r, 1/\mu_r)$, one can conclude that the TE and TM modes must swap upon the sign flip of the polarization indices (7): $v_1 \rightarrow -v_1$ or $v_2 \rightarrow -v_2$. This results in the diagram Fig. 3(b), where the lines $\varepsilon_r = \mu_r$ and $\varepsilon_r \mu_r = 1$ divide the $(\varepsilon_r, \mu_r)$-plane into alternating zones of TE and TM polarizations.

Note that according to the helicity-winding diagram Fig. 3(a), *both* TE and TM surface waves exist at every point of the "two-mode" zone $(\varepsilon_r < 0, \mu_r < 0)$, but only one of these modes is shown in Fig. 3(b). Showing both of these modes results in the two diagrams in Fig. 3(c), but only the first diagram corresponds to the *propagating* surface modes. Indeed, direct calculations show that the wavevectors of the surface modes have the form $k_\text{surf} \propto \sqrt{v_1 v_2}$ and $k_\text{surf} \propto \sqrt{-v_1 v_2}$ for the TM and TE polarizations, respectively (see Supplementary Materials). Hence one of these is always *real* (propagating modes in the first diagram Fig. 3(c)) while the other one is *imaginary* (evanescent surface modes in the second diagram Fig. 3(c)). Although these evanescent surface modes have never been considered before, these are observable, e.g., in the near-field scattering of surface electromagnetic waves. Furthermore, $k_\text{surf} \rightarrow 0$ for both propagating and evanescent surface modes at interfaces involving "epsilon-near-zero" or "mu-near-zero" materials, where the contribution of evanescent surface modes can become crucial.

Finally, there is one more feature in the characterization of surface modes, which is *not* determined by the topological invariants (3) and polarizations indices (7). Up to now, we have allowed any frequencies $\omega_\text{surf}$ of surface modes; however, because of the non-Hermitian character of the problem, these can also be either *real* or *imaginary*. In fact, the zones with real and imaginary frequencies are separated by the lines $\varepsilon_r = -1$ and $\mu_r = -1$, which correspond to plasmon resonances for a planar interface (see Supplementary Materials). Furthermore, since we reduced the non-Hermitian interface problem (5) and (6) to the problem with relative parameters $(\varepsilon_r, \mu_r)$ *and substitution* $\omega \rightarrow n_1 \omega$, as indicated in Eq. (8), the real-frequency and imaginary-frequency zones must swap upon the substitution $n_1^2 \rightarrow -n_1^2$. This splits the phase diagram of Fig. 3(c) into two diagrams for the $n_1^2 > 0$ and $n_1^2 < 0$ cases, as shown in Fig. 3(d). Considering only propagating surface modes with real $\omega_\text{surf}$ and $k_\text{surf}$, we find that the phase diagrams of Fig. 3(d) exactly coincide with rather sophisticated diagrams previously obtained in [33–35] by directly solving Maxwell equations. Importantly, the imaginary-$k$ and imaginary-$\omega$ surface waves were ignored in the previous studies, which resulted in truncated phase diagrams. Taking these modes into account makes the picture *complete* and *fully consistent* with the simple diagram in Figs. 3(a) described by the topological helicity winding number (3) and bulk-boundary correspondence (4).

## 3. Conclusions

We have shown that surface Maxwell waves have a fundamental topological origin which is described by the *helicity winding number* ($\mathbb{Z}_4$ or a pair of $\mathbb{Z}_2$ numbers) and *bulk-boundary correspondence*, Eqs. (3) and (4). On the one hand, the underlying mechanism resembles the $\mathbb{Z}_2$ winding number for the Dirac topological insulators with opposite-mass interfaces [1,2]. On the other hand, the situation is fundamentally different because we deal with the winding of the *helicity spectrum*, $\mathfrak{S}$, rather than that of the energy spectrum $E$. Moreover, the helicity operator in a medium is essentially *non-Hermitian*, and its spectrum can be either real or imaginary in lossless media. The helicity winding number labels the four topologically different quadrants of the parameter $(\varepsilon,\mu)$ space, which are separated by the *exceptional points* $\varepsilon = 0$ and $\mu = 0$ of the helicity-like Maxwell operator (5). In terms of momentum-space quantities, the non-Hermitian helicity leads to *complex Chern numbers* of photons in a medium, and their *phase* rather than the magnitude (as in the Hermitian case) corresponds to the helicity winding number.

The difference of the helicity winding number (3) between two media describes the number of surface Maxwell waves at the interface. Using different representations of the non-Hermitian helicity-based form of Maxwell equations, we introduce an additional pair of polarization $\mathbb{Z}_2$ indices (7). These are *not* topological numbers, they do not affect the number of surface modes, but these indices describe the *separation of the TE and TM polarizations* in the phase diagrams of surface modes, Fig. 3. Indeed, linking the polarization indices (7) to the phase diagram involves a spatial symmetry between the two media, which is broken when we replace the planar interface with a curved interface. In contrast, true topological phenomena are expected to survive at interfaces that break crystallographic symmetries.

Notably, due to their non-Hermitian origin, surface Maxwell waves are also essentially non-Hermitian modes. This means that these can have either real or imaginary frequencies and/or wave numbers. All previous studies of surface Maxwell waves considered only *propagating* surface waves with real parameters. This resulted in rather sophisticated and truncated phase diagrams [33–35]. Our theory augments this diagram with *evanescent surface modes* with imaginary parameters, which results in a very simple fundamental phase diagram Fig. 3(a) described by the topological helicity winding number (3).

Our theory provides new twists to several areas of wave physics: Maxwell electromagnetism, topological insulators, non-Hermitian quantum mechanics, and metamaterials. It shows that topological surface modes have been known and observed in electromagnetism long before the formulation of topological properties (e.g., surface plasmon-polaritons [14,15]). Furthermore, interfaces between the "negative-index" and "positive-index" metals provide "electromagnetic topological insulators" with no propagating bulk modes and topologically-protected surface modes [17,44]. We have also shown that macroscopic Maxwell equations in the helicity-based form naturally possess exceptional points in the $(\varepsilon,\mu)$-space and "chiral" non-Hermitian bulk modes in these points, which correspond to the "epsilon- and mu-near-zero" materials [17,40]. Finally, we note that our approach can be applied to other wave equations, providing an efficient model for systems described by non-Hermitian massless wave equations with helical bulk modes.

**Acknowledgements:** We are grateful to Filippo Alpeggiani and Ilya Shadrivov for fruitful discussions. This work was partially supported by the MURI Center for Dynamic Magneto-Optics via the Air Force Office of Scientific Research (AFOSR) (FA9550-14-1-0040), Army Research Office (ARO) (Grant No. Grant No. W911NF-18-1-0358), Asian Office of Aerospace Research and Development (AOARD) (Grant No. FA2386-18-1-4045), Japan Science and Technology Agency (JST) (Q-LEAP program, ImPACT program, and CREST Grant No. JPMJCR1676), Japan Society for the Promotion of Science (JSPS) (JSPS- RFBR Grant No. 17-52-50023, and JSPS-FWO Grant No. VS.059.18N), RIKEN-AIST Challenge Research Fund, the John Templeton Foundation, the Institute for Basic Science in Korea (IBS-R024-Y1), and the Australian Research Council.